\documentclass[twocolumn,nofootinbib]{revtex4}

\begin{document}

\title{Elementary charge and neutrino's mass from Planck length}

\author{Saulo Carneiro}

\affiliation{Instituto de F\'{\i}sica Gleb Wataghin, UNICAMP, 13083-970, Campinas, SP, Brazil\\Instituto de F\'{\i}sica, Universidade Federal da Bahia, 40210-340, Salvador, Bahia, Brazil}

\date{\today}

\begin{abstract}
It is shown that the postulation of a minimum length for the horizons of a black hole leads to lower bounds for the electric charges and magnetic moments of elementary particles. If the minimum length has the order of the Planck scale, these bounds are given, respectively, by the electronic charge and by $\mu \sim 10^{-21} \mu_B$. The latter implies that the masses of fundamental particles are bounded above by the Planck mass, and that the smallest non-zero neutrino mass is $m_{\nu} \sim 10^{-2}$eV. A precise estimation in agreement to the area quantisation of Loop Quantum Gravity predicts a mass for the lightest massive state in concordance with flavor oscillation measurements, and a Barbero-Immirzi parameter in accordance to horizon entropy estimations.
\end{abstract}

\maketitle

\section{Holography and Dirac coincidences}

Eight decades have passed since the appearance of the Dirac large number coincidences \cite{Dirac,Eddington}, and they still call our attention and reveal new facets. Among equivalent formulations, they can be expressed by verifying that the ratio between the mass of the observed universe $m_U$ and a typical baryon mass $m_N$ is related to the ratio between the latter and the Planck mass $m_P$, which by its turn is related to the ratio of gravitational to electrostatic interactions between two charged baryons. This is expressed mathematically  by
\begin{equation} \label{coincidence}
\left( \frac{m_U}{m_N} \right)^{1/4} \sim \frac{m_P}{m_N} \sim \frac{e}{\sqrt{G}m_N} \equiv \Omega,
\end{equation}
where $e$ and $G$ are respectively the electronic charge and the gravitational constant, and $\Omega \sim 10^{20}$. A possible explanation for the first coincidence comes from the so called holographic principle applied to cosmological scales \cite{mena,mena2}. This note discusses the second one. Concerning the first, let us assume that the entropy of the observable universe is bounded by $N \sim l_U^2/l_P^2$, as stated by the holographic principle \cite{thooft,holo1,holo3,holo4,holo5,holo6,holo7,holo8,holo9,holo10}. Here, $l_U = c/H$ is the Hubble-Lema\^itre horizon scale in the de Sitter limit, which for a spatially flat spacetime is given by $l_U = 2G m_U/c^2$, and the above bound can be rewritten as $N \sim m_U^2/m_P^2$. On the other hand, the maximum number of effective degrees of freedom in the observable volume is $N \sim l_U^3/l_N^3$, where $l_N = \hbar/(m_N c)$ is their wavelength \cite{mena,mena2}. Equating the two bounds above, it is straightforward to arrive at the first coincidence in (\ref{coincidence}).

\section{The elementary charge}

From (\ref{coincidence}) we have $m_P \sim \Omega m_N$, and from the value of the fine structure constant and the definition of the Planck mass it follows that $e \sim \sqrt{G} m_P$. From these two relations we obtain $\sqrt{G} m_N/e \sim \Omega^{-1}$, that is, the remaining coincidence we want to understand. Therefore, all we need is to explain the value of the fine structure constant.
As we will argue below, assuming the existence of a fundamental area $l_P^2$ as stated by the holographic conjecture, a quantum of charge emerges, given by $e \sim \sqrt{G} m_P$ as we need, since charges smaller than $e$ could otherwise form horizons with length smaller than $l_P$. Although a complete treatment of quantum black holes is still lacking (but see e.g. \cite{carr,gambini,perez,ashtekar}), this reasoning assumes that $l_P$ defines a fundamental scale for any degree of freedom. Let us initially consider the head-on scattering of two charges $e$ at a centre-of-mass energy $E$ high enough to form a black hole of gravitational radius $r_g$. The formed black hole has a mass $M \sim E/c^2 \sim e^2/(r_g c^2)$. Using $r_g = 2GM/c^2$, we have $e \sim \sqrt{G} M$. Introducing now the fine structure constant $\alpha = e^2/(\hbar c)$ and the Planck mass $m_{P} = \sqrt{\hbar c/G}$, we obtain
$M \sim \sqrt{\alpha}\, m_{P}$.
Therefore, the order of magnitude of the fine structure constant is related to the fact that the black hole mass cannot be smaller than the Planck mass. Note that the charges rest masses do not matter here, as they are negligible compared to $M$. The same result arises if we consider a Reissner-Nordstr\"om black hole \cite{BH} with $r_g/2 \geq r_Q \geq l_P$. Setting $r_g/2 \geq l_P$ leads to $M \geq m_P$. Taking $r_Q \equiv e\sqrt{G}/c^2 \geq l_P$, it follows that $e^2/(\hbar c) \geq 1$.  In this way, our argument is the same to say that the characteristic length scale $r_Q$ of the classical solution cannot be smaller than $l_P$.

\section{Magnetic moments and neutrino's mass}

If we now consider two parallel magnetic dipoles $\mu$ brought at a relative distance $r_g$, their interaction energy is given by $U = \mu^2/r_g^3$. Equating this energy to $Mc^2$ and postulating that $M \geq m_{P}$, we have
\begin{equation}\label{dipoles}
\mu \geq \frac{2\sqrt{2}\,c^2 l_P^2}{\sqrt{G}}.
\end{equation}
A similar result can be derived from the Kerr-Newman classical solution for a charged black hole with angular momentum $J$ \cite{BH}, by letting $r_g/2 \geq a \geq l_p$, where $a \equiv J/(Mc)$. This leads again to $M \geq m_P$ and, in addition, to $J \geq \hbar$, the quantum of action. Furthermore, the black hole magnetic moment is $\mu = e J/(Mc)$, and using $a \geq l_p$ we have
\begin{equation} \label{dipoles'}
\mu \geq \frac{e\hbar}{m_Pc},
\end{equation}
which differs from condition (\ref{dipoles}) by a factor of order $\sqrt{\alpha}$.
On the other hand, the magnetic moment of an elementary particle of mass $m$ and charge $e$ is typically given by $\mu \sim e \hbar/(mc)$. Hence, from (\ref{dipoles'}) we obtain the constraint
$m \leq m_P$,
which suggests that the Plank mass is an upper limit for the mass of any fundamental particle. From the combined constraints $m \leq m_P \leq M$ we may also infer that elementary particles cannot exhibit classical horizons.
Using $l_P =\sqrt{\hbar G/c^3}$, condition (\ref{dipoles}) reads
\begin{equation}\label{dipoles2}
\mu \geq 2.77 \times 10^{-21} \mu_B,
\end{equation} 
where $\mu_B = e \hbar/(2m_ec)$ is the Bohr magneton, with $m_e$ giving the electron mass. Among the neutral elementary particles, it is theoretically established that Dirac neutrinos carry magnetic moment, provided they have mass, as flavor oscillation measurements have been robustly indicating. The one-loop contributions to the neutrino magnetic dipole lead (in natural units) to\footnote{This is valid for Dirac neutrinos in the minimally extended Standard Model with right-handed singlets. Majorana neutrinos do not have magnetic moments.} \cite{dipole}
\begin{eqnarray}\label{dipoles4}
\mu_{\nu} &\approx& \frac{3eG_Fm_{\nu}}{8\sqrt{2}\pi^2} \\ &\approx& 3.20 \times 10^{-19} \left( \frac{m_{\nu}}{1\text{eV}} \right) \mu_B, \nonumber
\end{eqnarray}
where $G_F$ is the Fermi coupling constant and $m_{\nu}$ is the neutrino mass.
From (\ref{dipoles}) and (\ref{dipoles4}) we obtain for $m_{\nu}$ the lower bound\footnote{The last digit in this figure is affected by higher order corrections to (\ref{dipoles4}) that depend on the neutrinos mixing angles and Dirac phase \cite{dipole}. Using the current best-fits for these quantities \cite{neutrinos}, we find $m_{\nu} \approx 8.662\,(8) \times 10^{-3}\, \text{eV}$. Note, however, that higher order loops lead to corrections of the same order.}
\begin{eqnarray}\label{previsto}
m_{\nu} &\approx& \frac{32\pi^2}{3eG_Fm_P} \\ &\approx& 8.654 \times 10^{-3}\, \text{eV}. \nonumber
\end{eqnarray}
Incidentally, if one neutrino state is massless, oscillation experiments give for the lightest massive state (assuming normal ordering) the $1\sigma$  confidence interval \cite{neutrinos}
\begin{equation} \label{medido}
m_2 \approx (8.66 \pm 0.10) \times 10^{-3}\, \text{eV}.
\end{equation}

\section{A Quantum Gravity signature?}

Prediction (\ref{previsto}) and the observed value (\ref{medido}) differ by less than $0.1\%$. If not a coincidence, this would be a possible observational signature of spacetime quantisation. We do not have such a precision in the charge estimation of Section II, where the effects of vacuum polarisation were not taken into account. In the neutrino mass estimation, on the other hand, those are second order effects as the neutrino magnetic moment is itself an effect of vacuum fluctuations. We should also remark that the lower limit (\ref{dipoles}) is based on an extrapolation of the classical realm to the Planck scale. Although a lower bound $m_P$ for the black hole mass may be inferred from semi-classical reasonings \cite{carr}, it should be verified within a full quantum theory of gravity. This can be done for instance in the context of Loop Quantum Gravity \cite{lewandowski,thiemann}, that predicts the quantisation of horizons \cite{gambini,perez,ashtekar}. The unitary angular momentum of our {\it gedanken} neutrinos black hole can be written as the identity
\begin{equation} \label{J}
J = \frac{\sqrt{3}}{3} \sum_{i=1}^2 \sqrt{j_i(j_i+1)},
\end{equation}
with a $j_i = 1/2$ for each of the two parallel neutrinos. The event horizon radius $r_H$ is minimal in the extremal case $r_H^2 = a^2 = J$. Reminding that the horizon area is ${\cal A} = 4 \pi (r_H^2 + a^2)$, using (\ref{J}) we can write\footnote{We are using natural units $\hbar = c = G = 1$, in which $l_P = 1$.}
\begin{equation} \label{LQG}
{\cal A}/2 = 8\pi \gamma l_P^2 \sum_{i=1}^2 \sqrt{j_i(j_i+1)}.
\end{equation}
The right-hand side of Eq. (\ref{LQG}) can be identified with the eigenvalues of the LQG area operator\footnote{A general correspondence between angular momenta and spin network lines is conjectured in \cite{krasnov,bojowald}.} \cite{smolin}, with a Barbero-Immirzi parameter $\gamma = \sqrt{3}/6$, in $95\%$ agreement to the approximate value derived from the Bekenstein-Hawking entropy in the limit of large horizons \cite{meissner,mitra,corichi,agullo}. ${\cal A}/2$ is the effective area orthogonally pierced by the lines of a spin network with axial symmetry, given by the area of the equatorial circle ${\cal A}/4$ multiplied by $2$ punctures per line. In other words, the horizon is pierced by two lines of colour $1/2$, each line pierces the horizon twice, we then have four punctures, each one contributing with an area gap $4\sqrt{3} \pi \gamma l_P^2$, which totals a horizon area ${\cal A} = 16\sqrt{3} \pi \gamma l_P^2$, as given by (\ref{LQG}). It corresponds to a black hole mass $M = m_P$, as assumed in the previous section. As the black hole is extremal, there is no Hawking radiation and the horizon is isolated.

The same analysis can be made in the case of a charged, non-rotating black hole. Again, the horizon radius is minimal in the extremal limit, when $r_H = r_Q = M$ and there is no Hawking radiation. The horizon area is given by ${\cal A} = 4\pi r_H^2 = 4 \pi e^2$. On the other hand, in the fundamental representation of the spin network, the horizon isolation implies that the minimum eigenstate of the area operator has two punctures with quantum labels $(j_1,m_1) = (1/2,1/2)$ and $(j_2,m_2) = (1/2,-1/2)$, as this simultaneously gives the minimum area eigenvalue and fulfills the projection constraint \cite{meissner,mitra,corichi,agullo} 
\begin{equation}
\sum_i m_i = 0.
\end{equation}
Therefore, we now have
\begin{equation} \label{LQG2}
{\cal A} = 8\pi \gamma l_P^2 \sum_{i=1}^2 \sqrt{j_i(j_i+1)},
\end{equation}
which leads to $e^2 = 2\sqrt{3}\gamma$. For $\gamma  = \sqrt{3}/6$ we obtain $e = 1$ as in Section II. As already mentioned, it is worthy of note that we can determine the neutrinos mass state $m_2$ with $0.1\%$ error, while for the fundamental charge we have just a rough estimation, an order of magnitude above the observed value. This is in more accordance, by the way, to the original spirit of this paper, for which the orders of magnitude are what really matters. Nevertheless, if we do not consider the former precision as a mere coincidence, an explanation for the contrast with the latter should be outlined. Without expecting to give a definite answer, the difference may lie on the effects of vacuum polarisation on electric charges and magnetic moments. Loops corrections to the electron charge are logarithmically dependent on the energy scale, and the convergence of their series is not guaranteed at scales as high as the Planck energy. In fact, the above result $e = 1$ suggests that perturbative expansions cannot be used at such scales. In contrast, high order corrections to magnetic moments involve only powers of $\alpha/\pi \sim 10^{-3}$, where $\alpha$ is the low-energy fine structure constant \cite{mandl}. Any scale dependence is absorbed, after renormalisation, in the term $e G_F m_{\nu}$ of Eq. (\ref{dipoles4}), that has dimension of charge/mass. The perturbative expansion may be valid provided that further, quantum gravity corrections are negligible above the Planck length. This was also tacitly assumed when the classical expression for the black hole horizon area was used above.

\section*{Acknowledgements}

I am thankful to G. A. Mena Marug\'an for a critical reading and helpful suggestions. My thanks also to J.C. Fabris, R. Gambini, P.C. de Holanda, J. Olmedo, O.L.G. Peres, C. Pigozzo, A. Saa, R. Woodard and J. Zanelli for useful discussions. Work partially supported by CNPq.

\end{document}